\documentclass[twocolumn,prd,preprintnumbers,amsmath,amssymb]{revtex4}

\usepackage{graphicx}
\usepackage{dcolumn}
\usepackage{bm}
\usepackage{epsfig}
\usepackage{amssymb}
\usepackage{amsmath}

 \def\tskip{\setlength{\tskip}{5pt}}
\def\colwidth{\setlength{\colwidth}{3.5in}}

\newcommand{\lsim}{\mathrel{\hbox{\rlap{\lower.55ex\hbox{$\sim$}} \kern-.3em \raise.4ex \hbox{$<$}}}}
\newcommand{\gsim}{\mathrel{\hbox{\rlap{\lower.55ex\hbox{$\sim$}} \kern-.3em \raise.4ex \hbox{$>$}}}}
\newcommand{\be}{\begin{equation}}
\newcommand{\ee}{\end{equation}}
\newcommand{\ba}{\begin{eqnarray}}
\newcommand{\ea}{\end{eqnarray}}

\begin{document}

\title{Detecting relic gravitational waves in the CMB: A statistical bias}

\author{Wen Zhao}
\email{wzhao7@mail.ustc.edu.cn}\affiliation{Niels Bohr Institute,
Copenhagen University, Blegdamsvej 17, Copenhagen DK-2100,
Denmark}

\date{\today}

\begin{abstract}

Analyzing the imprint of  relic gravitational waves (RGWs) on the cosmic microwave background (CMB) power spectra provides a way to determine the signal of RGWs. In this Letter, we discuss a statistical bias, which could exist in the data analysis and has the tendency to overlook the RGWs. We also explain why this bias exists, and how to avoid it.

\end{abstract}


\maketitle


\section{Introduction \label{s-introduction}}

A stochastic background of relic gravitational waves was produced
in the very early stage of the Universe due to the superadiabatic
amplification of zero point quantum fluctuations of the
gravitational field \cite{grishchuk,star}. The relic gravitational
waves have a wide range of spreading of the spectra, and their
detection provides a direct way to study the physics in the early
Universe.

Recently, there have been several experimental efforts to
constrain the amplitude of relic gravitational waves in different
frequencies. Among various direct observations, LIGO S5 has
experimentally obtained so far the most stringent bound
$\Omega_{\rm gw}(f)\le6.9\times10^{-6}$ around $f\sim 100$Hz
\cite{ligo}, which will be much improved by future observations,
including the third-generation Einstein Telescope \cite{et}. The
timing studies on the millisecond pulsars by the PPTA and EPTA
teams also reported upper limits $\Omega_{\rm gw}(f)\lesssim
10^{-8}$ at $f\sim 1/{\rm yr}$ \cite{ppta,epta}. In addition,
there are two bounds on the integration $\int \Omega_{\rm
gw}(f)d\ln f\lesssim 1.5\times 10^{-5}$, obtained by the big bang
nucleosynthesis observation \cite{bbn} and the cosmic microwave
background radiation observation \cite{cmb3}.

In this paper, we shall focus on the detection of relic
gravitational waves by the cosmic microwave background (CMB)
radiation observations. The RGWs leave well understood imprints on
the anisotropies in temperature and polarization of CMB
\cite{cmb,cmb2}. The theoretical analysis of these imprints along
with the data (including $T$, $C$, $E$, $B$) from CMB experiments
allows one to determine the RGW background by constraining the
parameters: the tensor-to-scalar ratio $r$ and the spectral index
$n_t$. The current observations of CMB by WMAP satellite place an
interesting bound $r\le 0.20$ \cite{wmap7} by assuming $n_t=-r/8$,
which has been generalized in \cite{zhao2010}. These bounds are
equivalent to the constraints on the energy density $\Omega_{\rm
gw}(f)$ of relic gravitational waves at the lowest frequency range
$f\sim10^{-17}$Hz.

Detecting the relic gravitational waves remains one of the most
important tasks for the upcoming CMB observations (see \cite{task}
for reviews). Due to the various large contaminations, in the near
future, we can only expect to detect a signal of RGWs in a
relative low signal-to-noise ratio ($S/N$). This result would
guide the far future detections.

As for the whole data analysis, we expect that, the maximum value
of the parameters in the posterior possibility density function
(pdf) is unbiased for the `true' values of the parameters, which
is auto-satisfied when the $S/N$ is high. However when $S/N$ is
low, the maximum values of the parameters sometimes lead to a
biased guide for the `true' values, which can be generated either
by some systematics or by the statistics, and should be avoided in
any data analysis.

In this Letter, we will point out that,  a statistical bias could
exist in the CMB data analysis for the detection of RGWs.  We also
explain why the bias does exist, and suggest the way to avoid it.


\section{The statistical bias\label{s-spectra}}

The primordial power spectrum of relic gravitational waves can be
simply described by the following power-law formula: \ba\label{pt}
P_t(k)=A_t(k_0)\left({k}/{k_0}\right)^{n_t}, \ea where $k_0$ is
the pivot wavenumber, which can be arbitrarily chosen. $A_t(k_0)$
is the amplitude of RGWs, and $n_t$ is the spectral index. The
value of $n_t$ is quite close to zero, predicted by the physical
models of the early Universe. As usual, we can define the
tensor-to-scalar ratio $r\equiv A_t(k_0)/A_s(k_0)$, where
$A_s(k_0)$ is the amplitude of the density perturbations.
Obviously, assuming $A_s(k_0)$ is known as in this Letter, $r$ is
just $A_t(k_0)$ normalized by the constant $A_s(k_0)$.

In order to discuss the statistical bias for the detection of RGWs
in the data analysis, let us simulate the observable data for the
Planck satellite, where we only consider the Planck instrumental
noises at the $143$GHz frequency channels \cite{planck}. We adopt
the `input' cosmological models as $\Omega_b h^2=0.02267$,
$\Omega_ch^2=0.1131$, $\Omega_{\Lambda}=0.726$, $\tau_{\rm
reion}=0.084$, $h=0.705$, $A_s=2.445\times 10^{-9}$ and $n_s=1$.
The RGWs parameters are adopted as $r=\hat{r}=0.05$,
$n_t=\hat{n}_t=0$. As we have discussed in the previous paper
\cite{zhao}, this small $r$ is expected to be detected at
$2\sigma$ for the assumed noise level.

Based on this input cosmological model, and the assumed noise
level, we simulate $500$ data samples. For every sample, we can
probe the likelihood function by applying the Markov Chain Monte
Carlo (MCMC) method.  In the data analysis, we assume all the
parameters, except for $r$ and $n_t$, are all fixed as their input
values. For the parameter $n_t$, one always presumes the relation
$n_t=n_s-1$ or $n_t=-r/8$ in the data analysis
\cite{zbg}\cite{wmap5}. However, this assumption does depend on
the special cosmological models. If they are not the truth, but
presumed, the finial conclusion of the data analysis would deviate
from the real physics.

In order to avoid this danger, the natural way is setting $r$ and
$n_t$ as free parameters. We choose the flat priors of them in the
range $r\in[0,1]$ and $n_t\in [-3,3]$. We adopt the best-pivot
wavenumber, which is $k_0=0.0006$Mpc$^{-1}$ for the input model
and the assumed noise level \cite{zb}.

\begin{figure}[t]
\centerline{\includegraphics[width=10cm,height=6cm]{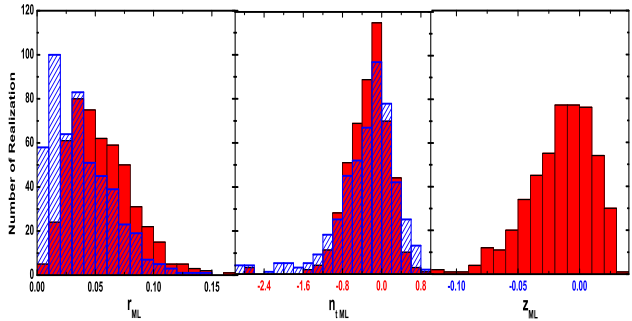}}
\caption{The distribution of $r_{\rm ML}$, $n_{t{\rm ML}}$ and $z_{\rm ML}$ for the $500$ simulated samples. The blue shadow shows the results by adopting the free parameters $r$ and $n_t$ and the flat prior of them. The red column shows the results by adopting the free parameters $r$ and $z$ and the flat prior of them.}\label{figure1}
\end{figure}

The most interesting final result is the maximum value in the
1-dimensional posterior pdf for the parameters $r$ and $n_t$. In
this paper, we denote them by $r_{\rm ML}$ and $n_{t{\rm ML}}$. Of
course, their values do depend on the simulated data. For
different data samples, they have different values. We expect the
distribution of these $500$ $r_{\rm ML}$ and $n_{t{\rm ML}}$ are
around their input values. However, it may be not the truth in the
real analysis. In Fig.\ref{figure1}, we plot the distribution of
$r_{\rm ML}$ and $n_{t{\rm ML}}$ with blue shadows. This figure
shows that, the distribution of $n_{t{\rm ML}}$ is peaked at zero,
the input value. However, the distribution of $r_{\rm ML}$
obviously approaches to $r=0$, and biased the input value at
$r=0.05$. This suggests that, if we deal with the data analysis in
this way, the resulting conclusion has the tendency to deviate
from the `true' value of $r$, and to overlook the RGWs.

\section{Understanding the statistical bias}
It is important to understand why this statistical bias does
exist. In order to realize it, let us proceed the following
analytical approximation for the likelihood analysis.

The primordial power spectrum of RGW in (\ref{pt}) can be
rewritten as, \ba\label{pt2}
P_t(k)=A_t(k_0)\left({k}/{k_0}\right)^{n_t}=A_s(k_0)r\exp\left[n_t\ln\left({k}/{k_0}\right)\right],
\ea which can be approximated as \ba\label{pt3} P_t(k)\simeq
A_s(k_0)\left[r+rn_t\ln\left({k}/{k_0}\right)\right]. \ea In this
approximation, we have used $|n_t|\ll 1$.

The total CMB power spectra $C_{\ell}^{Y}$ ($Y=T,C,E,B$) include
the contributions of density perturbations and gravitational
waves, i.e. \ba\label{cly} C_{\ell}^Y=C_{\ell,s}^Y+C_{\ell,t}^Y,
\ea where $C_{\ell,s}^Y$ and $C_{\ell,t}^Y$ are the contributions
of density perturbations and gravitational waves, separately. Note
that $C_{\ell,s}^{B}=0$. By considering $|n_t|\ll1$, the spectra
$C_{\ell,t}^Y$, as a function of $r$ and $n_t$, can be
approximated as \cite{zb} \ba\label{cly2} C_{\ell,t}^Y\simeq
\mathcal{C}_{\ell,t}^{Y}\left[r+rn_t\ln\left({\ell}/{\ell_0}\right)\right].
\ea Here $\mathcal{C}_{\ell,t}^{Y}\equiv
C_{\ell,t}^{Y}(r=1,n_t=0)$, and best-pivot multipole
$\ell_0=k_0\times 10^{4}$Mpc \cite{zb}. So $P_t(k)$ and
$C_{\ell,t}^{Y}$ are all the linear combinations of the parameters
$r$ and $rn_t$.

Now, let us turn to the likelihood function. The exact form can be
found in the previous works \cite{lewis}\cite{zbg,zhao,zb}. In the
analytical approximation, it can be well approximated by \cite{zb}
\ba\label{likelihood}
-2\ln\mathcal{L}=\sum_{\ell}\sum_{Y}\left(\frac{D_{\ell}^Y-C_{\ell}^Y}{\hat{\sigma}_{D_{\ell}^{Y}}}\right)^2.
\ea $D_{\ell}^{Y}$ is the observable data, and
$\hat{\sigma}_{D_{\ell}^{Y}}$ is standard deviation of
$D_{\ell}^{Y}$. The likelihood function (\ref{likelihood}) can be
rewritten as \cite{zb} \ba\label{likelihood2}
-2\ln\mathcal{L}=\sum_{\ell}\sum_{Y}\left[d_{\ell}^Y-(r+rn_tb_{\ell})a_{\ell}^Y\right]^2
\ea where we have defined the quantities \ba\label{dab}
d_{\ell}^Y\equiv\frac{D_{\ell}^Y-C_{\ell,s}^Y}{\hat{\sigma}_{D_{\ell}^{Y}}},~~a_{\ell}^Y\equiv\frac{\mathcal{C}_{\ell,t}^{Y}}{\hat{\sigma}_{D_{\ell}^{Y}}},~~b_{\ell}^{Y}\equiv\ln(\ell/\ell_0),
\ea which are all independent of the variables $r$ and $n_t$.
Obviously, the value of $d_{\ell}^{Y}$ depends on the data. For a
larger number of different sample, the average value of
$d_{\ell}^{Y}$ is $\langle
d_{\ell}^{Y}\rangle=a_{\ell}^{Y}(\hat{r}+\hat{r}\hat{n}_tb_{\ell})$,
due to the facts of $\langle
D_{\ell}^{Y}\rangle=C_{\ell,s}^{Y}+{C}_{\ell,t}^Y(r=\hat{r},n_t=\hat{n}_t)$
and ${C}_{\ell,t}^Y(r=\hat{r},n_t=\hat{n}_t)\simeq
\mathcal{C}_{\ell}^Y(\hat{r}+\hat{r}\hat{n}_tb_{\ell})$.

Since we have adopted the best-pivot multipole $\ell_0$, which is
defined by requiring \cite{zb}
$\sum_{\ell}\sum_{Y}(a_{\ell}^Y)^2b_{\ell}=0$, the likelihood
(\ref{likelihood2}) can be rewritten as \cite{zb}
\ba\label{likelihood3}
-2\ln\mathcal{L}=\left(\frac{r-r_p}{r_s}\right)^2+\left(\frac{rn_t-z_p}{z_s}\right)^2+C,
\ea where $C$ is a constant, and the other quantities are defined
by \ba\label{definitions}
r_p\equiv\frac{\sum_{\ell}\sum_{Y}a_{\ell}^Yd_{\ell}^Y}{\sum_{\ell}\sum_{Y}(a_{\ell}^Y)^2},~~
r_s\equiv\frac{1}{\sqrt{\sum_{\ell}\sum_{Y}(a_{\ell}^Y)^2}}, \\
\label{definitions2}
z_{p}\equiv\frac{\sum_{\ell}\sum_{Y}a_{\ell}^Yd_{\ell}^Yb_{\ell}}{\sum_{\ell}\sum_{Y}(a_{\ell}^Y
b_{\ell})^2},~~
z_{s}\equiv\frac{1}{\sqrt{\sum_{\ell}\sum_{Y}(a_{\ell}^Y
b_{\ell})^2}}. \ea

The posterior pdf relates to the likelihood by the prior. Here, let us adopt the flat prior for the parameters $r$ and $n_t$, the 2-dimensional posterior pdf for the variables is
\ba\label{prnt}
-2\ln P(r,n_t)=\left(\frac{r-r_p}{r_s}\right)^2+\left(\frac{rn_t-z_p}{z_s}\right)^2,
\ea
which follows the 1-dimensional posterior pdf for $r$ as follows,
\ba\label{pr}
P(r)=\frac{1}{r}\exp\left[-\frac{1}{2}\left(\frac{r-r_p}{r_s}\right)^2\right]+C'.
\ea

We notice that, when $r_p\gg r_s$, corresponding to $S/N\gg1$ (see
\cite{zb} for details), this pdf can be reduced that
\ba\label{pr2}
P(r)\simeq\frac{1}{r_p}\exp\left[-\frac{1}{2}\left(\frac{r-r_p}{r_s}\right)^2\right]+C'.
\ea This is gaussian function for $r$, and peaks at $r=r_p$ with
spread $r_s$. From the expression of $r_p$, we know that, the
value of $r_p$ depends on the data $D_{\ell}^Y$ by the quantity
$d_{\ell}^Y$. However, the average value of $r_p$ for a larger
number of sample is $\bar{r}_p=\hat{r}$, i.e. $r_p$ is an unbiased
estimator for $\hat{r}$. This has been mentioned in the previous
paper \cite{zb}.

But here, we want to emphasize that, when $r_p$ is not much larger
than $r_s$, the peak of the posterior pdf in (\ref{pr}) is smaller
than $r_p$, due to the term $1/r$. Especially when $r_p<3r_s$, the
peak of the pdf is very close to zero, which is never an unbiased
estimator for the input value $\hat{r}$. This explains what we
have found in the left panel of Fig.\ref{figure1}.

\section{Avoiding the statistical bias}
Now, let us consider the possible way to avoid this bias in the data analysis. Let us return to the likelihood function in (\ref{likelihood3}). We find that, if considering $r$ and $z\equiv rn_t$ as two independent parameters, this likelihood is a simple gaussian function for the uncorrected  parameters $r$ and $z$.

Now, we adopt the flat prior for the variables $r$ and $z$, and
the posterior pdf for $r$ and $z$ becomes \ba\label{prz} -2\ln
P(r,z)=\left(\frac{r-r_p}{r_s}\right)^2+\left(\frac{z-z_p}{z_s}\right)^2,
\ea from which follows that the 1-dimensional posterior pdf for
$r$ is \ba\label{[r3}
P(r)=\exp\left[-\frac{1}{2}\left(\frac{r-r_p}{r_s}\right)^2\right]+C'.
\ea This pdf peaks at $r=r_p$, which is an unbiased estimator for
the input value $\hat{r}$. Similarly, we can also find that, the
1-dimensional posterior pdf for $z$ peaks at $z=z_p$, which is
also an unbiased estimator for $\hat{z}\equiv\hat{r}\hat{n_t}$. So
the statistical bias in data analysis is elegantly avoided.

In order to clearly show this result, we have analyzed the same
$500$ samples, by adopting the flat prior on $r$ and $z$. In
Fig.\ref{figure1}, we plot the distribution of the $r_{\rm ML}$
and $z_{\rm ML}$ with the solid columns. As expected, we find
that, these $r_{\rm ML}$ and $z_{\rm ML}$ are all distributed
around at their input values $\hat{r}=0.05$ and $\hat{z}=0$, and
the bias for the tensor-to-scalar ratio is naturally avoided. In
this figure, we also plot the distribution of $n_{t{\rm ML}}$,
which also unbiased distributed around its input value
$\hat{n}_t=0$.

It is interesting to compare the difference between the prior $f(r,z)$ and the general prior $f(r,n_t)$. They can be related by the Jacobi, i.e.
\ba\label{jacobi}
f(r,n_t)=\left|\frac{\partial(r,z)}{\partial(r,n_t)}\right|f(r,z)=rf(r,z).
\ea
This relation shows that, the flat prior $f(r,z)=1$ exactly corresponds to $f(r,n_t)=r$. So, comparing with the analysis with flat prior $f(r,n_t)=1$, the new flat prior $f(r,z)$ induces a larger value of the variable $r$.

~

\section{conclusion}
In this Letter, we find a statistical bias in the CMB data
analysis for the detection of RGWs, when the signal-to-noise ratio
is not very high. This could overlook the signal of RGWs in the
CMB data analysis. We explain why this bias does exist by the
analytical approximation of the likelihood function, and also find
this bias can be elegantly avoided by adopting the orthogonalized
parameters $r$ and $z\equiv rn_t$, instead of the general
parameters $r$ and $n_t$.

In the end, we should emphasize that a similar statistical bias
might exist for any data analysis \cite{prior}, which should be
carefully treated.



~

{\bf Acknowledgement:} The author thanks D.Baskaran, L.P.
Grishchuk, P.Coles,  H.Chiaka, S.Gupta for helpful discussions.
This work is supported by NSFC grants Nos. 10703005, 10775119 and
11075141.


\begin{thebibliography}{35}

\bibitem{grishchuk}
L.P.~Grishchuk, Zh.~Eksp.~Teor.~Fiz.~{\bf 67}, 825 (1974);
Ann.~N.~Y.~Acad.~Sci~{\bf 302}, 439 (1977); Pis'ma
Zh.~Eksp.~Teor.~Fiz.~{\bf 23}, 326 (1976); Uspekhi Fiz.~Nauk {\bf
121}, 629 (1977).


\bibitem{star}
A. A. Starobinsky, JETP Lett. {\bf 30}, 682 (1979); Phys. Lett. B
{\bf 91}, 99 (1980).


\bibitem{ligo}
B. P. Abbott {\it et al.,}  (LIGO Scientific Collaboration and
Virgo Collaboration), Nature (London) {\bf 460}, 990 (2009).


\bibitem{et}
S. Hild {\it et al.,} arXiv:1012.0908; C. Van Den Broeck {\it et
al.,} private communication.



\bibitem{ppta}
F. Jenet {\it et al.,} Astrophys. J. {\bf 653}, 1571 (2006).


\bibitem{epta}
R. van Haasteren {\it et al.,} arXiv:1103.0576; W. Zhao, Phys. Rev
D {\bf 83}, 104021 (2011).


\bibitem{bbn}
B. Allen and J. D. Romano, Phys. Rev. D {\bf 59}, 102001 (1999).

\bibitem{cmb3}
T. L. Smith, E. Pierpaoli and M. Kamionkowski, Phys. Rev. Lett.
{\bf 97}, 021301 (2006).


\bibitem{cmb}
U. Seljak and M. Zaldarriaga, Phys. Rev. Lett. {\bf 78}, 2054
(1997); M. Kamionkowski, A. Kosowsky and A. Stebbins, Phys. Rev.
Lett. {\bf 78}, 2058 (1997).


\bibitem{cmb2}
J. R. Printchard and M. Kamionkowski, Ann. Phys. (N.Y.) {\bf 318},
2 (2005); W. Zhao and Y. Zhang, Phys. Rev. D {\bf 74}, 083006
(2006); D. Baskaran, L. P. Grishchuk and A. G. Polnarev, Phys. Rev
D {\bf 74}, 083008 (2006).




\bibitem{wmap7}
E. Komatsu {\it et al.,} Astrophys. J. Suppl. Ser. {\bf 192}, 18
(2011).

\bibitem{zhao2010}
W. Zhao and L. P. Grishchuk, Phys. Rev. D {\bf 82}, 123008 (2010).





\bibitem{task}
J.  Bock, {\it et al.,} astro-ph/0604101; D. Baumann {\it et al.,}
AIP Conf. Proc. {\bf 1141}, 10  (2009); COrE Collaboration,
arXiv:1102.2181.








\bibitem{planck}
Planck Collaboration, arXiv:astro-ph/0604069.

\bibitem{zhao}
W. Zhao, Phys. Rev. D {\bf 79}, 063003 (2009).

\bibitem{zbg}
W. Zhao, D. Baskaran and L.P. Grishchuk, Phys. Rev. D {\bf 79}, 023002 (2009), Phys. Rev. D {\bf 80}, 083005 (2009).

\bibitem{wmap5}
E. Komatsu {\it et al}., Astrophys. J. Suppl. Ser. {\bf 180}, 330 (2009).




\bibitem{zb}
W. Zhao and D. Baskaran, Phys. Rev. D {\bf 79}, 083003 (2009).


\bibitem{lewis}
S. Hamimeche and A. Lewis, Phys. Rev. D {\bf 77}, 103013 (2008).

\bibitem{prior}
e.g. W. Valkenburg, L.M. Krauss and J. Hamann, Phys. Rev. D {\bf 78}, 063521 (2008).


\end{thebibliography}
\end{document}